\documentclass[conference,a4paper]{IEEEtran}
\IEEEoverridecommandlockouts
\usepackage[ansinew]{inputenc}
\usepackage[T1]{fontenc}
\usepackage{color,amsmath,enumerate,amssymb,hhline,verbatim}
\usepackage{url}
\usepackage{multirow}
\usepackage{textcomp}
\usepackage{algorithm}
\usepackage[noend]{algpseudocode}

\newcommand{\pars}{(n,k,d,r,\delta)}
\newcommand{\Z}{\mathcal{Z}}
\newcommand{\cl}{\mathrm{cl}}
\newcommand{\te}[1]{\textrm{#1}}
\newcommand{\q}{\textquotesingle}
\newcommand{\kr}{\left\lceil k/r \right\rceil}
\newcommand{\lkr}{\left\lceil k/r \right\rceil}

\newcommand{\bblue}{\begin{color}{blue}}
\newcommand{\eblue}{\end{color}}

\newcommand{\bred}{\begin{color}{red}}
\newcommand{\ered}{\end{color}}

\begin{document}

\title{Bounds on the Maximal Minimum Distance of Linear Locally Repairable Codes }



\author{
\IEEEauthorblockN{Antti P\"oll\"anen\IEEEauthorrefmark{1}, Thomas Westerb\"ack\IEEEauthorrefmark{1}, Ragnar Freij-Hollanti\IEEEauthorrefmark{2}, and Camilla Hollanti\IEEEauthorrefmark{1}}
\IEEEauthorblockA{
\IEEEauthorrefmark{1}Department of Mathematics and Systems Analysis, Aalto University, P.O.Box 11100, FI-00076 Aalto, Finland\\
Emails: \{firstname.lastname\}@aalto.fi}
\IEEEauthorblockA{\IEEEauthorrefmark{2}Department of Communications and Networking, Aalto University, P.O.Box 13000, FI-00076 Aalto, Finland\\
Email: ragnar.freij@aalto.fi}
\thanks{This work was partially supported by the Academy of Finland grants \#276031, \#282938, \#283262, and by Magnus Ehrnrooth Foundation, Finland. The support from the European Science Foundation under the COST Action IC1104 is also gratefully acknowledged. The first author would like to thank the Department of Mathematics and Systems Analysis at Aalto University for  financial support during the summer of 2015.}
}



\maketitle

\newtheorem{definition}{Definition}[section]
\newtheorem{thm}{Theorem}[section]
\newtheorem{proposition}[thm]{Proposition}
\newtheorem{lemma}[thm]{Lemma}
\newtheorem{corollary}[thm]{Corollary}
\newtheorem{exam}{Example}[section]
\newtheorem{conj}{Conjecture}
\newtheorem{remark}{Remark}[section]
\newtheorem{set_construction}{A construction of matroids}[section]

\begin{abstract}
Locally repairable codes (LRCs) are error correcting codes used in distributed data storage. Besides a global level, they enable errors to be corrected locally, reducing the need for communication between storage nodes. There is a close connection between almost affine LRCs and matroid theory which can be utilized to construct good LRCs and derive bounds on their performance.

A generalized Singleton bound for linear LRCs with parameters $(n,k,d,r,\delta)$ was given in 
 [N. Prakash \textit{et al.}, ``Optimal Linear Codes with a Local-Error-Correction Property'', IEEE Int. Symp. Inf. Theory]. In this paper, a LRC achieving this bound is called \emph{perfect}. Results on the existence and nonexistence of linear perfect $(n,k,d,r,\delta)$-LRCs were given in [W. Song \textit{et al.}, ``Optimal locally repairable codes'', IEEE J. Sel. Areas Comm.]. Using matroid theory, these existence and nonexistence results were later strengthened in [T.~Westerb\"ack \textit{et al.}, ``On the Combinatorics of Locally Repairable Codes'', Arxiv: 1501.00153], which also provided a  general lower bound on the maximal achievable minimum distance $d_{\rm{max}}(n,k,r,\delta)$ that a linear LRC with parameters $(n,k,r,\delta)$ can have.  This article expands the class of parameters $(n,k,d,r,\delta)$ for which there exist perfect linear LRCs and improves the lower bound for $d_{\rm{max}}(n,k,r,\delta)$. Further, this bound is proved to be optimal for the class of matroids that is used to derive the existence bounds of linear LRCs.

\end{abstract}

\section{Introduction}

In modern times, the need for large scale data storage is swiftly increasing. This need is present for example in large data centers and in cloud storage. The large scale of these distributed data storage systems makes hardware failures common. However, the data should be preserved regardless of failures, and error correcting codes can be utilized to prevent data loss.

A traditional approach is to look for codes which simultaneously maximize error tolerance and minimize storage space consumption. However, this tends to yield codes for which error correction requires an unrealistic amount of communication between storage nodes. \textit{Locally repairable codes} (LRCs) solve this problem by allowing errors to be corrected locally, in addition to the global level. 

Besides the parameters $(n,k,d)$ referring to the length, dimension, and minimum distance of a regular linear code, respectively, a LRC is characterized by two additional parameters, $r$ and $\delta$. Informally speaking, the local error correction is enabled by dividing the code symbols into locality sets whose size is at most $r+\delta-1$ and inside which any $\delta-1$ symbols can be recovered using the rest of the symbols in the locality set. 

\subsection{Related Work}

The notion of a LRC was first introduced in \cite{papailiopoulos2014locally}. The generalized Singleton bound for linear $(n,k,d,r,\delta)$-LRCs states that 
\begin{equation} \label{eq:singleton}
d \leq n - k + 1 - \left ( \left\lceil k/r \right\rceil - 1 \right)(\delta - 1).
\end{equation}
This bound was given in \cite{gopalan12} for $\delta=2$ and in \cite{prakash2012optimal} for a general $\delta$. This bound has then been generalized for both linear and nonlinear codes in several ways, see \emph{e.g.}  \cite{LRCpapailiopoulos}, \cite{cadambe13}, \cite{rawat15} and \cite{tamo16}. 

The class of \emph{almost affine} codes is a generalization of the class of linear codes. In \cite{simonis98} it was proved that every almost affine code induces a matroid. Many important properties (but not all) of almost affine codes are \emph{matroid invariants} in the sense that the properties only depend on the matroid structure of the code. Matroid theory was used in \cite{tamo2013optimal} in order to prove that the minimum distance of a class of linear LRCs achieves the generalized Singleton bound.  It was proved in \cite{westerback15} that every almost affine LRC induces a \emph{matroid} such that the parameters $(n,k,d,r,\delta)$ of the LRC appear as matroid invariants.
Consequently, the parameters $(n,k,d,r,\delta)$ were generalized to matroids and the bound \eqref{eq:singleton} was proven to also hold for all matroids, which is nontrivial since not all matroids are induced by almost affine codes. 
 An even more general Singleton bound was given for polymatroids in \cite{westerback2015applications}, motivated by the fact that all general LRCs induce a polymatroid.

Results on the existence and non-existence of linear $(n,k,d,r,\delta)$-LRCs achieving the generalized Singleton bound were given in \cite{song2014optimal}. Codes or matroids achieving the generalized Singleton bound are here called \emph{perfect}.  Using the \emph{lattice of cyclic flats} of matroids, the non-existence results of \cite{song2014optimal} were strengthened in  \cite{westerback15}.


There are many different constructions of perfect LRCs, e.g. see \cite{prakash2012optimal}, \cite{tamo2013optimal}, \cite{song2014optimal} \cite{silberstein2013optimal}, \cite{tamo2014family}. Using a matroid-based construction in \cite{westerback15}, classes of linear LRCs with a large span on the parameters $(n,k,d,r,\delta)$ and local repair sets were given. By this construction, linear perfect $(n,k,d,r,\delta)$-LRCs were constructed for all the parameters from the existence results given in \cite{song2014optimal}. Further, again by the matroid-based construction, a general lower bound was given on the maximal achievable minimum distance $d_{\rm{max}}(n,k,r,\delta)$ that a linear LRC with parameters $(n,k,r,\delta)$ can have.

\subsection{Contributions}

This paper strengthens several results given in \cite{westerback15}. Firstly, using the matroid-based construction we extend the class of linear perfect $(n,k,d,r,\delta)$-LRCs with $\lkr=2$. Secondly, we improve the general lower bound on $d_{\rm{max}}(n,k,r,\delta)$ for linear LRCs and prove that the new bound is optimal for the matroid-based construction. The results of this paper were originally presented in the bachelor thesis of the first author \cite{kandi}, which provides a more comprehensive account as well as full proofs.

\section{Preliminaries} \label{sec:polymatroids-codes}

\subsection{Almost Affine Locally repairable codes}

In this section, we will define an almost affine $(n,k,d,r,\delta)$-LRC. As usual, $n$ denotes the length of a codeword and $d$ its minimum (Hamming) distance. An almost affine code is defined as follows: 

\begin{definition}
	A code $C \subseteq \Sigma^n$, where $\Sigma$ is a finite set of size $s \geq 2$, is \emph{almost affine} if for each $X \subseteq [n]$ we have $\log_s(|C_X|) \in \mathbb{Z}$. 
\end{definition}

Here $[n]=\{1,2,...,n\}$ and $C_X$ denotes the projection of the code $C$ to $\Sigma^{|X|}$, \emph{i.e.}, $C_X = \{(c_{i_1},...,c_{i_{m}}): \mathbf{c} = (c_1,...,c_n) \in C\}$, where $X = \{i_1, \ldots, i_m\} \subseteq [n]$. 
The parameter $k$ is, as usual, defined as $k=\log_s(|C|)$.

The local error correction of a LRC is performed inside \emph{$(r,\delta)$-locality sets}: 

\begin{definition}
	When $1\leq r \leq k$ and $\delta \geq 2$, an $(r,\delta)$-locality set of $C$ is a subset $S\subseteq [n]$ such that
	\begin{equation*}
	\begin{alignedat}{3}
	& \textrm{(i)}  && |S|\leq r+\delta-1, \\
	&  \textrm{(ii)} \ && d(C_S) \geq \delta, \textrm{where $d(C_S)$ is the min. distance of $C_S$}.
	\end{alignedat}
	\end{equation*}
\end{definition}

We say that $C$ is a \emph{locally repairable code} with \emph{all-symbol locality} $(r,\delta)$ if every code symbol $l \in [n]$ is included in an $(r,\delta)$-locality set.

\subsection{Matroids}

Matroids are combinatorial structures that capture, in an abstract sense, a certain kind of dependence common to various mathematical structures. 
Of the numerous equivalent matroid definitions, we will use the one utilizing the \emph{rank function} $\rho$. In the following, $2^E$ denotes the set of all subsets of $E$. 

\begin{definition}
	\label{DefMat}
	A matroid $M=(E,\rho)$ is a finite set $E$ along with a rank function $\rho: 2^E \rightarrow \mathbb{Z}$ satisfying the following conditions for every subsets $X,Y \subseteq E$:
	\begin{equation*}
	\begin{alignedat}{3}
	& \textrm{(i)} \quad && 0 \leq \rho (X) \leq |X|, \\
	& \textrm{(ii)} \quad && X \subseteq Y \Rightarrow \rho (X) \leq \rho (Y), \\
	& \textrm{(iii)} \quad && \rho(X) + \rho(Y) \geq  \rho (X \cup Y) + \rho(X \cap Y).
	\end{alignedat}
	\end{equation*}
\end{definition}

This definition is for instance satisfied by the set of column vectors $E$ of a matrix over a field, and $\rho(X)$ being equal to the rank of the submatrix consisting of the column vectors indexed by $X$.
If $E$ is the set of edges of an undirected graph, then a matroid is obtained by letting $\rho(X)$ be the size of a minimal spanning tree of the subgraph with edges $X$. 


Next, we define some matroid concepts relevant to us.
 A subset $X\subseteq E$ is said to be \emph{independent} if $\rho(X)=|X|$. The \emph{nullity} of a set $X \subseteq E$ is defined by $\eta(X) = |X|-\rho(X)$. 

A \emph{circuit} is a dependent set $X \subseteq E$ whose all proper subsets are independent, \emph{i.e.},  $\rho(X \setminus \{x\}) = \rho(X) = |X|-1 $ for every $x \in X$. A set $X\subseteq E$ is $cyclic$ if it is a union of circuits. We denote the sets of circuits and cyclic sets of a matroid by $\mathcal{C}(M)$ and $\mathcal{U}(M)$, respectively.

The \emph{closure} of a set $X \subseteq E$ is defined by $\cl(X) = \{x \in E: \rho(X \cup \{x\}) = \rho(X)\}$. A set $X\subseteq E$ is a \emph{flat} if $X=\cl(X)$. A \emph{cyclic flat} is a flat that also is a cyclic set.

The \textit{restriction of $M=(E,\rho)$ to $X$} is the matroid $M|X=(X,\rho_{|X})$ where $\rho_{|X}(Y)=\rho(Y)$ for $Y \subseteq X$. 



A \textit{lattice} is a partially ordered set for which every pair of two elements has a unique infimum, \textit{meet}, and a unique  supremum, \textit{join}. The cyclic flats of a matroid have the property that they form a finite lattice $(\mathcal{Z}, \subseteq)$ with meet $X\wedge Y = \bigcup_{C \in \mathcal{C}(M): C \subseteq X\cap Y} C$ and join $X \vee Y = \cl(X \cup Y)$, for $X,Y \in \mathcal{Z}$ \cite{bonin2008lattice}.

The least element of the lattice is the element $0_{\mathcal{Z}}\in \mathcal{Z}$ such that  $X \subseteq 0_{\mathcal{Z}} \Rightarrow X=0_{\mathcal{Z}}$ for every $X\in \mathcal{Z}$. Correspondingly, the greatest element   is the element $1_{\mathcal{Z}}\in \mathcal{Z}$ such that $1_{\mathcal{Z}} \subseteq X \Rightarrow X=0_{\mathcal{Z}}$ for every $X\in \mathcal{Z}$. 

The sets of the atoms $A_{\mathcal{Z}}$ and coatoms $coA_{\mathcal{Z}}$ are defined by 
$A_{\mathcal{Z}}=\{X\in \mathcal{Z} \setminus \{0_{\mathcal{Z}}\}: \nexists Y \in \mathcal{Z}\te{ such that }  0_{\mathcal{Z}} \subsetneq Y \subsetneq X\}$ and $coA_{\mathcal{Z}}=\{X\in \mathcal{Z} \setminus \{1_{\mathcal{Z}}\}: \nexists Y \in \mathcal{Z}\te{ such that }  X \subsetneq Y \subsetneq 1_{\mathcal{Z}} \}$, respectively.

Matroids can also be defined via this \emph{lattice of cyclic flats}, which is our main tool for constructing and analyzing matroids in this paper. The associated axioms are presented in the following theorem:
\begin{thm}[\cite{bonin2008lattice}]
	Let $\mathcal{Z} \subseteq 2^E$ and let $\rho$ be a function $\rho: \mathcal{Z}\rightarrow \mathbb{Z}$. There is a matroid $M$ on $E$ for which $\mathcal{Z}$ is the set of cyclic flats and $\rho$ is the rank function restricted to the sets in $\mathcal{Z}$ if and only if 
	\begin{equation*}
	\begin{alignedat}{2}
	(Z0) \quad & \te{$\mathcal{Z}$ is a lattice under inclusion,} \\
	(Z1) \quad & \rho(0_{\mathcal{Z}})= 0, \\
	(Z2) \quad & X,Y \in \mathcal{Z} \te{ and } X \subsetneq Y \Rightarrow \\
	& 0 < \rho(Y)-\rho(X)<|Y|-|X|,\\
	(Z3) \quad & X,Y \in \mathcal{Z} \Rightarrow\rho(X) + \rho(Y) \geq \\
	& \rho(X \vee Y) + \rho(X \wedge Y) + |(X \cap Y) \setminus (X \wedge Y)|.
	\end{alignedat}
	\end{equation*}
\end{thm}


\section{Matroids and LRCs}

\subsection{Relationship between matroids and almost affine LRCs}

The following theorem defines the associated matroid $M_C$ of an almost affine code $C$.  

\begin{thm}[\cite{simonis98}]
Let $C \subseteq \sum^n$ be an almost affine code, where $|\sum| = s$. Then $M_C = ([n], \rho_C)$ is a matroid, where \vspace{-0.2 cm}
$$
\rho_C(X) = \log_s(|C_X|) \hbox{, for } X \subseteq [n]. \vspace{-0.2 cm}
$$
\end{thm}
The following result can be viewed as a definition of the parameters $(n,k,d,r,\delta)$ for a matroid from the viewpoint of its cyclic flats. Hence, the parameters $(n,k,d,r,\delta)$ of an almost affine LRC $C$ can be analyzed using its associated matroid $M_C = (\rho_C, [n])$ in the theorem below.

	\begin{thm}[\cite{westerback15}]
		Let $M=(E,\rho)$ be a matroid with \newline $0 < \rho(E)$ and $1_{\mathcal{Z}} = E$. Then \vspace{-0.2 cm}
		\label{thmParamsFromCyclicFlats}
		\begin{equation*}
		\begin{alignedat}{3}
		&(i) \quad && n=|1_{\mathcal{Z}}|, \\
		&(ii) \quad && k = \rho(1_{\mathcal{Z}}), \\
		&(iii) \quad && d=n-k+1-\max\{\eta(Z): Z \in coA_{\mathcal{Z}}\}, \\
		&(iv) \quad && \te{$M$ has locality $(r,\delta)$ if and only if for each $x \in E$} \\
		& && \te{there exists a cyclic set $S_x \in \mathcal{U}(M)$ such that} \\
		& && \te{a) } x \in S_x, \\
		& && \te{b) } |S_x| \leq r+\delta-1, \\
		& && \te{c) } d(M|S_x) = \\
		& && \quad \eta(S_x)+1-\max\{\eta(Z): Z \in coA_{\mathcal{Z}(M|S_x)}\} \geq \delta.
		\end{alignedat}
		\end{equation*}
		\end{thm}

\subsection{Matroid-based constructions of linear LRCs}

The matroid-based construction of linear LRCs that is used in the constructive proofs of both \cite{westerback15} and this article is the following:

	\emph{Construction 1} \cite{westerback15}: Let $F_1,...,F_m$ be a collection of subsets of a finite set $E$, $k$ a positive integer, and $\rho: \{F_i\}_{i \in [m]} \rightarrow \mathbb{Z}$ a function such that
	\begin{equation}
	\label{eqCorollIII.1}
	\begin{alignedat}{3}
	& (i) \quad && 0<\rho(F_i)< |F_i| \te{ for } i \in [m], \\
	& (ii) \quad && F_{[m]} = E, \\
	& (iii) \quad && k \leq F_{[m]}-\sum_{i \in [m]} \eta(F_i), \\
	& (iv) \quad &&  |F_{[m]\setminus \{j\}} \cap F_j|<\rho(F_j) \te{ for all } j \in [m],
	\end{alignedat}
	\end{equation}
		where for every element $i \in [m]$ and subset $I \subseteq [m]$, \vspace{-0.1 cm}
	\begin{equation*}
	\begin{alignedat}{3}
	& \te{(a) \quad}&&\eta(F_i)=|F_i|-\rho(F_i)\,, \\
	& \te{(b) \quad}&& F_I=\bigcup_{i \in I} F_i\,.
	\end{alignedat}\vspace{-0.3 cm}
	\end{equation*}
	
	Further, we extend $\rho$ to a function for subsets $I \subseteq [m]$ by \vspace{- 0.2 cm}
	
	$$\rho(F_I) = \min\{|F_{I}|-\sum_{i \in I} \eta(F_i), k \}\,.$$

	\vspace{- 0.2 cm}

\begin{thm}[\cite{westerback15}]
		\label{thmc1props}
	The previous construction defines a matroid $M(F_1,...,F_m;k;\rho)$ which equals $M_C=([n],\rho_C)$ for some linear LRC $C$ over a sufficiently large $\mathbb{F}_q$ such that 	

		$$
		\begin{alignedat}{3}
			& \te{(i)} \quad && \Z = \{F_I:I\subseteq [m], \rho(F_I)<k  \} \cup E\,, \\
			& \te{(ii)} && n=|E|\,, \\
			& \te{(iii)} && k=\rho(E)\,, \\
			& \te{(iv)} && d = n-k+1 -\max\{\sum_{i \in I}\eta(F_i): F_I \in \Z \setminus E \}\,, \\
			& \te{(v)} && \delta-1 = \min_{i \in [m]}\{\eta(F_i)\}\,, \\
			& \te{(vi)} && r = \max_{i \in [m]}\{\rho(F_i)\}\,.
		\end{alignedat}
		$$
		
		For each $i \in [m]$, any subset $S \subseteq F_i$ with $|S|=\rho(F_i)+ \delta-1$ is a locality set of the matroid.
		
	\end{thm}

The motivation to use this construction comes from the fact that a matroid from it has a maximal $d$, given the matroid's  set of atoms $\{F_i\}$, rank function $\rho: \{F_i\} \rightarrow \mathbb{Z}$ restricted to the atoms, and dimension $k$. This follows from the fact that its cyclic flats $F_I$ have minimal size and maximal rank, achieving the bound in Z3 when $\rho(F_I)<k$. 
		

In a proof given later, we will use the following more specialized version of the matroid-based construction given above.

			\emph{Graph construction 1: (\cite[v2]{westerback15})} Let $G = G(\alpha, \beta, \gamma; k, r, \delta)$ be a graph with vertices $[m]$ and edges $W$, where $(\alpha, \beta)$ are two functions $[m] \rightarrow \mathbb{Z}$, $\gamma: W \rightarrow \mathbb{Z}$, and $(k,r,\delta)$ are three integers with $0<r<k$ and $\delta \geq 2$, such that \vspace{-0.1 cm}
			\begin{equation}
			\label{eqconds1}
			\begin{alignedat}{2}
			& \textrm{(i)} \quad \textrm{$G$ is a graph with no 3-cycles,} \\
			& \textrm{(ii)} \quad  0 \leq \alpha (i) \leq r-1 \textrm{ for } i \in [m], \\
			& \textrm{(iii)} \quad  \beta(i) \geq 0 \textrm{ for } i \in [m], \\
			& \textrm{(iv)} \quad  \gamma(w) \geq 1 \textrm{ for } w \in W, \\
			& \textrm{(v)} \quad  k \leq rm - \sum_{i \in [m]} \alpha(i) - \sum_{w \in W} \gamma(w), \\
			& \textrm{(vi)} \quad r - \alpha(i) > \sum_{w=\{i,j\} \in W} \gamma(w) \textrm{ for } i \in [m].
			\end{alignedat} \vspace{-0.1 cm}
			\end{equation}

			\begin{thm}[\cite{westerback15}, v2]
				\label{thmConstr1}
				Let $G(\alpha, \beta, \gamma; k, r, \delta)$ be a graph on $[m]$ such that the conditions (i)-(vi) given in \eqref{eqconds1} are satisfied. Then there is an $(n,k,r,d,\delta)$-matroid $M(F_1,...,F_m;k;\rho)$ given by Theorem \ref{thmc1props} with \vspace{-0.5 cm}
				
				\begin{align*}
				\begin{split}
				& \textrm{(i)} \quad n = (r+\delta-1)m - \sum_{i \in [m]} \alpha(i) + \sum_{i \in [m]} \beta(i) - \sum_{w \in W} \gamma(w), \\
				& \textrm{(ii)} \quad d = n-k+1 - \max_{I \in V_{<k}} \{(\delta-1)|I|+\sum_{i \in I} \beta(i) \},
				\end{split}
				\end{align*}\vspace{-0.4 cm}
				
				where
				\begin{equation*}
				V_{<k} = \{I \subseteq [m]: r|I|- \sum_{i \in I} \alpha(i) - \sum_{i,j \in I, w = \{i,j\}\in W} \gamma(w) < k\}.
				\end{equation*}
				
			\end{thm}


\section{Main results}

Our first result is an expanded class of parameters $(n,k,r,\delta)$ for which the generalized Singleton bound \eqref{eq:singleton} can be achieved for linear LRCs. The previous bound in \cite{westerback15} was identical to this bound for $2a \leq r-1$ but weaker otherwise. The parameter restrictions $0<r<k\leq n - \kr(\delta-1)$ and $\delta \geq 2$ are required for $(n,k,d,r,\delta)$-matroids to exist \cite{westerback15}. \newline

\begin{thm}
	\label{thmResult1}
Define $a=r\lkr -k$ and $b=(r+\delta-1)\lceil\frac{n}{r+\delta-1}\rceil-n$, and let $(n,k,r,\delta)$ be integers such that $0<r<k\leq n - \kr(\delta-1)$, $\delta \geq 2$, $b>a \geq \lceil k/r \rceil -1$, and $\lceil k/r \rceil = 2$. If \vspace{- 0.2 cm}
	\begin{equation}
	\left\lceil \frac{n}{r+\delta-1} \right\rceil \geq \left\lceil b/a \right\rceil + 1\,,
	\end{equation}
	then the maximal achievable minimum distance for linear LRCs with parameters $(n,k,r,\delta)$ is 
	 $$d_{\rm{max}} = n-k+1-\left( \kr -1\right)(\delta-1)\,.$$
	
	\begin{IEEEproof}  We prove our result by giving an explicit construction of perfect matroids $M(F_1,\ldots,F_m;k;\rho)$ of Thm. \ref{thmc1props} for the desired parameter values.
		
		\emph{A matroid construction.} Let $n'$, $r'$, $\delta'$, and  $k$ be integers such that $0<r'<k\leq n' - \left\lceil k/r' \right\rceil(\delta'-1)$, $\delta' \geq 2$, $b'>a'$, and $m \geq \left\lceil b'/a' \right\rceil + 1$, where we define \vspace{-0.4 cm}
		
		\begin{align*}
		& b' = \left\lceil \frac{n'}{r'+\delta'-1} \right\rceil (r'+\delta'-1) - n' , \\
		& a'=\left\lceil k/r' \right\rceil r' -k ,\\
		& m = \left\lceil \frac{n'}{r'+\delta'-1} \right\rceil.
		\end{align*}
		
		Let $F_1, ..., F_m = \{F_i\}_{i \in [m]}$ be a collection of finite sets with $E = \bigcup_{i \in m}F_i$ and $X \subseteq E$ a set such that \vspace{- 0.2 cm}
		
		\begin{equation*}
		\begin{alignedat}{3}
		&\te{(i)} \quad && F_i \cap F_j \subseteq X \quad \te{ for }i,j \in [m] \te{ with } i \neq j, \\
		&\te{(ii)} \quad && |X| = a', \\
		&\te{(iii)} \quad && |F_i| = r' + \delta' - 1 \quad \te{ for } i \in [m], \\
		&\te{(iv)} \quad && |F_i \cap X| = a' \te{ for } 1 \leq i \leq \left\lceil b'/a' \right\rceil, \\
		&\te{(v)} \quad && |F_i \cap X| = b'- \left(\left\lceil b'/a' \right\rceil  - 1 \right)a' \te{ for } i = \left\lceil b'/a' \right\rceil + 1,\\
		&\te{(vi)} \quad && |F_i \cap X| = 0 \te{ for } i > \left\lceil b'/a' \right\rceil + 1. \\
		\end{alignedat}
		\end{equation*}
		
		Let $\rho$ be a function $\rho:   \{F_i\}_{i \in [m]} \rightarrow \mathbb{Z}$ such that $\rho(F_i)=r'$ for each $i \in [m]$.
		
		For the rest of the proof, we first check that this construction satisfies the conditions in \eqref{eqCorollIII.1}. Then we use Theorem \ref{thmc1props} to show that it yields perfect matroids (and thus linear LRCs) for the desired class of parameters $(n,k,r,\delta)$,  which are shown to equal their primed counterparts. The details of this can be found in \cite{kandi}.

	\end{IEEEproof}
\end{thm}

Our second main result is an improved lower bound for $d$. The actual improvement is the bound \eqref{newbound} as the bound \eqref{remainingbound} is identical to what was used in \cite{westerback15}. \newline

\begin{thm}
	\label{raja}
	Let $(n,k,r,\delta)$ be integers such that $0<r<k\leq n - \kr(\delta-1)$, $\delta \geq 2$, and $b>a$. Also let $m=\left\lceil \frac{n}{r+\delta-1} \right\rceil - 1$ and $v = r+\delta-1-b-\left\lfloor \frac{r+\delta-1-b}{m} \right\rfloor m$. 	Then for linear LRCs with parameters $(n,k,r,\delta)$:  
	
	If $\delta-1 \leq \left( \left\lceil k/r \right\rceil -1\right) \left\lfloor \frac{r+\delta-1-b}{m} \right\rfloor
	+ \min\{v, \kr -1\}$, we have
	\begin{equation}
	\label{remainingbound}
	d_{\rm{max}} \geq n-k+1-\kr(\delta-1).
	\end{equation}
	
	Otherwise, if $\delta-1 > \left( \left\lceil k/r \right\rceil -1\right) \left\lfloor \frac{r+\delta-1-b}{m} \right\rfloor
	+ \min\{v, \kr -1\}$, then \vspace{-0.4 cm}
	
	\begin{equation}
	\label{newbound}
	\begin{split}
	d_{\rm{max}} \geq \ & n-k+1 - \min\left\{v, \kr -1\right\}\\ 
	& - \left( \kr -1\right)\left(\left\lfloor \frac{r+\delta-1-b}{m} \right\rfloor+\delta-1\right).
	\end{split}
	\end{equation}

	
	We denote the right side of the bound \eqref{newbound} by $d_{\rm{new}}$. This bound is an improvement over its counterpart $d_{\rm{old}} = n-k+1-\kr(\delta-1)+(b-r)$ in \cite{westerback15}, since \vspace{- 0.3 cm}

	\begin{equation}
	\label{parannus}
	d_{\rm{new}} - d_{\rm{old}} \geq \left\lfloor \frac{r+\delta-1-b}{m} \right\rfloor \left(m-\left\lceil k/r \right\rceil +1\right) \geq 0.
	\end{equation}
	
\end{thm}

\begin{IEEEproof}
	Let $n' \in \mathbb{Z}$ be such that it satisfies the conditions for $n$ in Theorem \ref{raja}. 
	
	\emph{A graph construction}. Let $G(\alpha, \beta, \gamma; k, r, \delta)$ be intended as an instance of Graph construction 1 with \vspace{- 0.4 cm}
	
	\begin{align}
	\begin{split}
	& \textrm{(a)} \quad m = \left\lceil \frac{n'}{r+\delta-1} \right\rceil - 1,\\
	& \textrm{(b)} \quad  W=\emptyset, \\
	& \textrm{(c)} \quad \alpha (i)= 0 \textrm{ for } i \in [m], \\
	& \textrm{(d)} \quad \beta (i) = \begin{cases} \left\lceil \frac{r+\delta-1-b'}{m} \right\rceil \textrm{ for } 1 \leq i \leq v' \vspace{1.5 mm}, \\
	\left\lfloor \frac{r+\delta-1-b'}{m} \right\rfloor \textrm{ for } v' <  i \leq m,
	\end{cases}
	\end{split}
	\end{align}

	where $b' = \left\lceil \frac{n'}{r+\delta-1} \right\rceil (r+\delta-1) - n'$ and $v' = r+\delta-1-b'-\left\lfloor \frac{r+\delta-1-b'}{m} \right\rfloor m$.  
	
	The rest of the proof consists of checking that the conditions in \eqref{eqconds1} are satisfied and using Theorem \ref{thmConstr1} to show that the construction yields the expected $d$ for all desired parameter sets $(n,k,r,\delta)$. Finally, the inequalities in \eqref{parannus} will be proved. A full version of the proof can be found in \cite{kandi}.

\end{IEEEproof}

\begin{exam} To see that the difference $d_{\rm{new}}-d_{\rm{old}}$ is not identically zero, consider for instance the graph construction used in the proof with  parameter values $n'=139, k=60, r=20, \delta=21$.
\end{exam}

 Lastly, we show that the bound in Thm. \ref{raja} for matroids (linear LRCs) from Construction 1  is tight for parameter sets $(n,k,r,\delta)$ for which there exists no perfect matroid (linear LRC) from Construction 1. \newline 

\begin{thm}
	\label{thmResult3}
Let $(n,k,r,\delta)$ be integers such that there exists no perfect $(n,k,d',r,\delta)$-matroid from Construction 1.  Let $M$ be an $(n,k,d,r,\delta)$-matroid from Construction 1 and let us denote the bound in Theorem \ref{raja} by $d_{b}=d_{b}(n,k,r,\delta)$. Then $d \leq d_b$.

\end{thm}

\begin{IEEEproof}
	A more detailed proof can be found in \cite{kandi}. Let $M=M(F_1,...,F_m;k;\rho)$ be a matroid from Construction 1 for which there exists no perfect matroid from the same construction with the same parameters $(n,k,r,\delta)$.
	
	Assume that $\max\{|I| : F_I \in \mathcal{Z}_{<k}\} \geq \lkr$. Using Theorem \ref{thmc1props} (iii), we then obtain $ d \leq n-k+1-\kr(\delta-1)$, as $\eta(F_i) \geq \delta-1$ for every $i \in [m]$. 
	
	Thus the theorem holds in this case and we are only left with the case $\max\{|I| : F_I \in \mathcal{Z}_{<k}\} = \lkr - 1$, as we easily see that $\max\{|I| : F_I \in \mathcal{Z}_{<k}\}<\lkr-1$ is impossible.

	There must be an atom $F_i$ with $\eta(F_i) > \delta-1$, since otherwise the matroid would be perfect. Next we show that our current assumptions imply $m < \lceil \frac{n}{r+\delta-1} \rceil$. We do this by showing that $m \geq \lceil \frac{n}{r+\delta-1} \rceil$ would allow the existence of perfect matroids, which is a contradiction. The perfect matroids are constructed by, roughly speaking, repeatedly decreasing the nullity of atoms $F_u$ with $\eta(F_u)>\delta-1$ by an element of $F_u$ to another atom $F_i$. which either has $\rho(F_i)<r$ or overlaps with another atom $F_k$. In the former case, $\rho(F_i)$ will be increased by one, and in the latter case, the element in the intersection will no longer be part of $F_i$.

	Let us denote $s=\sum_{i \in [m]} \eta(F_i)$. Let us distribute this nullity evenly among the atoms $F_i$, \emph{i.e.}, set \vspace{- 0.4 cm}
	
\begin{equation*}
	\eta (F_i) = \begin{cases} \left\lceil s/m \right\rceil \textrm{ for } 1 \leq i \leq s- \left\lfloor s/m \right\rfloor m \vspace{1.5 mm}, \\
	\left\lfloor s/m \right\rfloor \textrm{ for } s- \left\lfloor s/m \right\rfloor m <  i \leq m.
	\end{cases}
	\end{equation*}
	
	For minimizing $\max\left\{\sum_{i \in I}\eta(F_i): |I|= \kr-1 \right\}$, this setup is clearly optimal and yields the bound \vspace{-0.4 cm}
	
	\begin{equation}
	\label{eqBoundTasajako}
	\begin{split}
	& \max\left\{\sum_{i \in I}\eta(F_i): |I|= \kr-1 \right\} \\
	& \geq \left(\kr-1\right)\left\lfloor s/m \right\rfloor + \min\left\{\kr-1, s- \left\lfloor s/m \right\rfloor m  \right\}.
	\end{split}
	\end{equation}
	
	
	The bound in \eqref{eqBoundTasajako} is clearly increasing as a function of $s$, and $s$ is bounded by $s \geq n-rm$. Thus we obtain the bound

	\vspace{- 0.3 cm}
	\begin{equation}
	\label{eqBoundTasajako2}
	\begin{split}
	& \max\left\{\sum_{i \in I}\eta(F_i): |I|= \kr-1 \right\} \geq \left(\kr-1\right)\left\lfloor \frac{n-rm}{m} \right\rfloor\\
	& \hspace{28pt}+ \min\left\{\kr-1, n-rm- \left\lfloor \frac{n-rm}{m} \right\rfloor m  \right\}.
	\end{split} \vspace{-0.5 cm}
	\end{equation}

	\vspace{-0.2 cm}This bound is in turn decreasing as a function of $m$ and we can obtain a new bound by substituting $m = \lceil \frac{n}{r+\delta-1} \rceil-1$. 
	By additionally substituting $v$ and $b$ by their definitions in \eqref{newbound}, we can see that the bounds \eqref{newbound} and \eqref{eqBoundTasajako2} are equal.
	
	We have thus proved that the value of $d$ for non-perfect matroids is always bounded from above by either the bound \eqref{remainingbound} or the bound \eqref{newbound}. This proves the theorem.

\end{IEEEproof}

\begin{remark}
 The class of matroids constructed in \eqref{eqCorollIII.1} constitutes a small subclass of the class of matroids called gammoids \cite{westerback15}. A method of constructing linear codes from  gammoids can be extracted by using \cite{lindstrom1973vector}. The smallest field size required by LRCs is an important issue, since it affects the computational complexity of the code. In general for gammoids there is a known upper bound for the field size, $2^n$ \cite{lindstrom1973vector}. However, we are convinced that this bound is not tight for the construction given in \eqref{eqCorollIII.1}. We have ongoing research on explicit constructions of linear LRCs over small fields obtained from \eqref{eqCorollIII.1} and conjecture an upper bound on the smallest  field size that is polynomial with $n$. However, explicit constructions of LRCs for the matroid-based construction given in \eqref{eqCorollIII.1} are out of the scope of this paper. 
\end{remark}

\section{Conclusions}

 In this paper, we provided an expanded class of parameters for which perfect linear LRCs exist (Thm. \ref{thmResult1}). We also gave a general lower bound for the maximal minimum distance $d_{max}$ (Thm. \ref{raja}), which we proved to be optimal for sub-perfect LRCs from Construction 1 (Thm. \ref{thmResult3}).


{}


These theorems suggest the following two-stage approach for solving $d_{max}(n,k,r,\delta)$ for almost affine LRCs: The first goal is to derive an expression for $d_{max}$ restricted to sub-perfect LRCs. 
Then, full knowledge of $d_{max}$ would be achieved by determining the class of parameters $(n,k,r,\delta)$ for which perfect LRCs exist.


Theorem \ref{thmResult3} is an attempt at accomplishing the first task. It is only a partial result towards this goal as it is limited to matroids from Construction 1. However, matroids from Construction 1 have a maximal $d$ given their setup of atoms, which suggests that the bound in Theorem \ref{raja} is tight or almost tight in the general case for sub-perfect matroids.

Theorem \ref{thmResult1} in turn is an addition to the existing results on for which parameter values perfect matroids exist. A complete solution of this second question would seem to require solving hard problems of extremal set theory.

\bibliography{LRCimprovements}{}

\begin{thebibliography}{10}

\bibitem{papailiopoulos2014locally}
D.~S. Papailiopoulos and A.~G. Dimakis, ``Locally repairable codes,'' {\em IEEE
  Trans. Inf. Theory}, vol.~60, no.~10, pp.~5843--5855, 2014.

\bibitem{gopalan12} P. Gopalan, C. Huang, H. Simitci, and S. Yekhanin,  ``On the locality of codeword symbols,'' \emph{IEEE Trans. Inf. Theory}, 58(11), pp. 6925--6934, 2012.

\bibitem{prakash2012optimal}
N.~Prakash, G.~M. Kamath, V.~Lalitha, and P.~V. Kumar, ``Optimal linear codes
  with a local-error-correction property,'' in {\em 2012 IEEE Int. Symp. Inf.
  Theory (ISIT)}, pp.~2776--2780.

\bibitem{LRCpapailiopoulos} D. S. Papailiopoulos, and A. G. Dimakis,  ``Locally repairable codes,'' \emph{2012 IEEE Int. Symp. Inf. Theory  (ISIT)}, pp. 2771--2775.

\bibitem{cadambe13} V. Cadambe and A. Mazumdar,
``An upper bound on the size of locally recoverable codes'', In \emph{Proc. IEEE Symp. Netw. Coding}, pp. 1--5, Jun. 2013.

\bibitem{rawat15} A. S. Rawat, A. Mazumdar and S. Vishwanath ``Cooperative local repair in distributed storage,'' EURASIP J. Adv. Sign. Proc, online 2015.

\bibitem{tamo16}
I. Tamo, A. Barg and A. Frolov, "Bounds on the parameters of locally recoverable codes", arXiv: 1506.07196.

\bibitem{simonis98} J. Simonis and A. Ashikhmin, ``Almost affine codes'', \emph{Design, codes and cryptography}, 14, pp. 179--197, 1998.

\bibitem{tamo2013optimal}
I.~Tamo, D.~S. Papailiopoulos, and A.~G. Dimakis, ``Optimal locally repairable
  codes and connections to matroid theory,'' in {\em 2013 IEEE Int. Symp. Inf.
  Theory (ISIT)}, pp.~1814--1818.

\bibitem{westerback15}
T.~Westerb\"ack, R.~Freij-Hollanti, T.~Ernvall, and C.~Hollanti, ``On the
  combinatorics of locally repairable codes via matroid theory.'' arXiv:
  1501.00153.
  
  \bibitem{westerback2015applications}
T.~Westerb\"ack, R.~Freij-Hollanti, and C.~Hollanti, ``Applications of
  polymatroid theory to distributed storage systems,'' {\em in proc. 53rd
  Annual Allerton Conf. on Comm. Control}, 2015.
  
\bibitem{song2014optimal}
W.~Song, S.~H. Dau, C.~Yuen, and T.~J. Li, ``Optimal locally repairable linear
  codes,'' {\em IEEE J. Sel. Areas Commun.}, vol.~32, no.~5, pp.~1019--1036,
  2014.
  
\bibitem{silberstein2013optimal}
N.~Silberstein, A.~S. Rawat, O.~O. Koyluoglu, and S.~Vishwanath, ``Optimal
  locally repairable codes via rank-metric codes,'' in {\em 2013 IEEE Int.
  Symp. Inf. Theory (ISIT)}, pp.~1819--1823.

\bibitem{tamo2014family}
I.~Tamo and A.~Barg, ``A family of optimal locally recoverable codes,'' in {\em
  IEEE Trans. Inf. Theory}, vol.~60, no.~8, pp.~4661--4676, 2014.

\bibitem{kandi}
A.~P\"oll\"anen, ``Locally repairable codes and matroid theory,'' bachelor
  thesis, Aalto University, arXiv: 1512.05325, 2015.







\bibitem{bonin2008lattice}
J.~E. Bonin and A.~De~Mier, ``The lattice of cyclic flats of a matroid,'' {\em
  Annals of Combinatorics}, vol.~12, no.~2, pp.~155--170, 2008.

\bibitem{lindstrom1973vector}
B.~Lindstr{\"o}m, ``On the vector representations of induced matroids,'' {\em
  Bull. London Math. Soc.}, vol.~5, no.~1, pp.~85--90,
  1973.

\end{thebibliography}
\bibliographystyle{ieeetr}


\end{document}